**Seasonality of primary productivity affects coastal species more than its magnitude**

Carlota Muñiz[1]*, Christopher D. McQuaid[1], Nicolas Weidberg[1,2]

[1] Department of Zoology and Entomology, Rhodes University, Grahamstown, 6140, South Africa.

[2] present address: Facultade de Ciencias do Mar, Universidade de Vigo, Vigo, 36200, Spain, and Department of Biological Sciences, University of South Carolina, Columbia, 29208, USA.

*Corresponding author; e-mail: carlota.fernandezmuniz@gmail.com

**Abstract**

While the importance of extreme conditions is recognised, patterns in species' abundances are often interpreted through average environmental conditions within their distributional range. For marine species with pelagic larvae, temperature and phytoplankton concentration are key variables. Along the south coast of South Africa, conspicuous spatial patterns in recruitment rates and the abundances of different mussel species exist, with focal areas characterized by large populations. We studied 15 years of sea surface temperature (SST) and chlorophyll-a (chl-a) satellite data, using spectral analyses to partition their temporal variability over ecologically relevant time periods, including seasonal (101 to 365 days) and intra-seasonal cycles (20 to 100 days). Adult cover and mussel recruitment were measured at 10 sites along the south coast and regression models showed that about 70% of the variability in recruitment and adult cover was explained by seasonal variability in chl-a, while mean annual chl-a and SST only explained 30% of the recruitment, with no significant effect for adult cover. SST and chl-a at two upwelling centres showed less predictable seasonal cycles during the second half of the study period with a significant cooling trend during austral autumn, coinciding with one of the mussel reproductive peaks. This likely reflects recent changes in the Agulhas Current, the world's largest western boundary current, which affects coastal ecosystems by driving upwelling. Similar mechanisms probably operate in other marine systems with the potential to affect the distribution patterns of key ecosystem engineers. We propose that variability in the characteristic timescales of environmental fluctuations can explain the spatial patterns of abundance of foundational species by affecting larval recruitment. This is especially important in a context of global and pervasive climate change, as shifts in the periodicity of environmental fluctuations appear to reflect large scale climatic teleconnections driven by anthropogenic forcing.

1. Introduction

Identifying the main environmental factors that drive the distributions of organisms and the spatial variability in their biological fitness is a central aim in ecological studies (Krebs 1985, Verberk 2011). Traditionally, it is thought that spatial variability in the energy balance of a species determines its reproductive output and in turn its abundance. Such energy balance in turn depends not only on food availability and gradients of environmental stress, but also on biological interactions like predation and competition (Hall et al. 1992, Seebacher and Franklin 2012). Theoretically, the outcome of trade-offs between energy investment and reproductive output is an abundant centre distribution with higher abundances at the core of the distributional range, where environmental conditions are optimal and impose little physiological stress on individuals (Andrewartha and Birch 1954). However, recent reviews found less support for the abundant centre hypothesis than expected (Sagarin and Gaines 2002, Dallas et al. 2017) and tests on several species across wide latitudinal ranges have shown different spatial patterns (e.g. Rivadeneira et al. 2010; Baldanzi et al. 2013).

A recognised problem with evaluating the abundant centre hypothesis or other spatial descriptions of abundance is the difficulty of sampling sufficient populations across the species' range (Sagarin and Gaines 2002), but another, overlooked issue is background variability in environmental conditions across that range. Not only can environmental conditions fail to follow a clear centre-to-edge gradient, but the gradients usually considered to be behind the observed organismal distributions are gradients of mean values (Potapova and Charles 2002, Lagos et al. 2008, Rivadeneira et al. 2010). Thus, little attention is paid to the temporal scales at which these variables fluctuate significantly, although many are known to vary in space as a function of latitude, elevation, or depth (Brock et al. 1993, Tapia et al. 2014, Hua 2016). Some time scales, such as annual seasonality, determine to a great extent biological phenology by setting synchronising environmental cues that are predictable in time, and which trigger key biological processes like migration, growth and reproduction (Williams et al. 2017). Despite the obvious importance of the temporal scales of environmental variability, these scales are rarely identified or used to explain the performance or distribution of species. Therefore, deviations from the abundant centre hypothesis consisting, for instance, of irregular distributions with local abundance hotspots could arise not only from irregularly distributed conditions, but also from spatial variations in the way those conditions change with time (e.g. Baldanzi et al. 2015). Thus, the intrinsic temporal scales at which environmental variables fluctuate in time may vary in space and be a better predictor of organismal distributions. Thus, the mechanisms behind irregular patterns cannot be inferred if the relative importance of different timescales is not even measured.

In the case of intertidal invertebrates with biphasic life cycles, a dispersive planktonic phase determines spatial connectivity within the metapopulation and complicates the effects of environmental variability. The life cycles of such species involve external fertilisation and the development of a planktonic larva that lives for weeks to months in the pelagic realm, feeding mainly on phytoplankton (de Schweinitz and Lutz 1976, Vargas et al. 2006, Carrier et al. 2018). Thus, predictable, seasonal peaks in primary productivity can have multiple beneficial effects on benthic populations, by promoting synchronous spawning (Starr et al. 1990), and providing food for larvae in the form of phytoplankton, as well as high quality food that can improve the condition and physiological state of adults (Newell and Bayne 1980, Norkko and Thrush 2006). Thus, although the effect of larval advection precludes a direct relationship between reproductive output and recruitment, we would expect a relationship at the metapopulation level.

Mytilid mussels have a biphasic life cycle and on the coasts of Africa, from West Africa to Moçambique, the native *Perna perna* and *Choromytilus meridionalis*, and the invasive *Mytilus galloprovincialis* provide examples of complex distributional patterns that do not conform to the abundant centre hypothesis (Harris et al. 1998). On the western, Atlantic, coast of South Africa, the strong, permanent upwelling characterising the Benguela Current system supports high primary production which in turn sustains large mussel populations and high recruitment of mussel larvae (Harris et al. 1998, Reaugh-Flower et al. 2011). Both primary production and mussel recruitment rates are notably lower on the south coast of South Africa, where upwelling is rare (Reaugh-Flower et al. 2011) yet recruitment rates and adult abundances at some sites resemble the high levels of the west coast (Bownes and McQuaid 2006, von der Meden et al. 2008). Critically, localised upwelling on the south coast can be driven by both wind and the dominant Agulhas Current (Gill and Schuman 1979, Leber et al. 2017), and influences nutrient and chlorophyll levels with quite different temporal variability from the west coast (Schumann et al. 1982).

Given ongoing anthropogenic global climate change, accounting for variability in key environmental factors is central to explaining distributional shifts in species (Malhi et al. 2010, Wang et al. 2015). In addition, climate change alters not only mean values, but also the dominant periodicities in key environmental determinants such as primary production, temperature and, in the ocean, pH (Henson et al. 2010, González Taboada and Anadón 2014, Kwiatkowski and Orr 2018, Santer et al. 2018). In the case of the south coast of South Africa, observed ongoing shifts in the circulation of the Agulhas Current within the study region and beyond (Beal and Elipot 2016) are likely to affect not only mean thermal and primary productivity conditions, but also their seasonality. To what extent these changes in the prevalent timescales of variability may affect populations is poorly understood.

Here, we highlight the role of the scales of temporal environmental variability in determining complex spatial distributions of species. To do this, we used the spatial patterns of mussel abundances and onshore recruitment rates along the Southern coast of South Africa as a model example to evaluate the biological relevance of both mean values and the significant timescales of variability of chlorophyll-a concentrations and sea surface temperature. Dominant periodicities in primary production and thermal conditions were quantified and examined over 15 years of the 21st century in order to identify the potential of temporal variability in large scale climatic forcing to drive distributional shifts.

**2. Materials and methods**

2.1. Spatial patterns of temporal variability in environmental variables

The study area extended along the south coast of South Africa, from its southernmost point (i.e. Cape Agulhas, 19.94°E) to 28.48°E (Fig. 1A). To cover the water mass likely to interact with or affect coastal communities, data were collected for the area from the coastline to approximately 120km offshore, thus covering most of the inner Agulhas Bank. Sea surface temperature (SST) and chlorophyll-a (chl-a) data were obtained from the Moderate Resolution Imaging Spectroradiometer (MODIS Aqua https://oceancolor.gsfc.nasa.gov/), with a resolution of 4km (4km blocks are referred to as *pixels*). Level-3 composite images of eight-day averages were obtained to maximise coverage and minimise data gaps for the period from the 9th July 2002 until the 27th September 2017 (Table 1). The 4x4km pixels were used to calculate an average value for 12x20km areas (across and alongshore respectively), from here on referred to as *megapixels*. Megapixels were used to simplify the analysis of such a large spatial and temporal dataset, and to reduce the high frequency of missing values in individual pixels due to cloud cover or quality control that are usually present in satellite derived measurements. Coastal waters are optically complex areas where chl-a estimates may not only be an estimate of primary productivity but may also reflect other coloured particles derived from sediments, or bottom reflection in shallow waters (Tilstone et al. 2013). Although these shortcomings are acknowledged, and chl-a measures are not a measure of primary production, for simplicity, chl-a is treated here as an estimate of food available for coastal organisms, in the form of phytoplankton or, to a lesser extent, other coloured dissolved organic matter, and treated as a proxy for primary production.

Spectral analysis was performed on each megapixel of the eight-day time series to determine the temporal frequencies which explained the most variability in SST and chl-a. Total variance was partitioned into four periods of interest that are likely to be important for ecological systems, following Tapia et al. (2014), and calculating the sum of the spectral densities within each period. Periods were defined as: (1) frequencies greater than one year and shorter than three years, which could be affected by climatic oscillations such as the Indian Ocean Dipole (384 - 1152 days, referred to as *inter-annual*), (2) frequencies close to an annual cycle which affect temperate systems (101 - 383 days, referred to as *seasonal*), (3) frequencies of variability shorter than the annual cycle, which could reflect variability induced by the Agulhas Current (20 - 100 days, i.e. *intra-seasonal*), (4) short-term frequencies of variability influenced by passing storms (16 - 19 days, *synoptic*). This produced the cumulative spectral densities for each period, which were used as a proxy of how much variability each respective period accounted for. Prior to analysis, linear interpolation was applied to remove missing values (missing data in chl-a and SST datasets, respectively: mean (±S.D.) = 4.79% (±4.11) and 1.19% (±0.75), maximum = 14.84% and 5.71%, longest gap = 40 and 16 days). Dates at the beginning or at the end of the series containing missing values after linear interpolation were removed. The chl-a dataset extended from the 25$^{th}$ July 2002 to the 11$^{th}$ September 2017 (total 697 eight-day composites), and the SST dataset from the 9$^{th}$ September 2002 to the 11$^{th}$ September 2017 (total 699 eight-day composites). Spectral densities were estimated for each megapixel using the function *spectrum*, from the package *stats* in R (R Core Team, 2015, version 3.2.0).

### 2.1.1. Changes in temporal variability during the period of study

In addition, the dataset used to examine the frequencies of variability of SST and chl-a was divided in two halves to determine if SST or chl-a showed a change of periodicity during the 15 years of data, indicating changes in the seasonality of either variable. To achieve this, data were divided into two blocks: 25$^{th}$ July 2002 – 15$^{th}$ February 2010 and 16$^{th}$ February 2010 – 11$^{th}$ September 2017 for chl-a, and 9$^{th}$ July 2002 – 7$^{th}$ February 2010 and 8$^{th}$ February 2010 – 27$^{th}$ September 2017 for SST. Due to the length of the dataset and the periods considered when calculating the frequencies of variability (i.e., inter-annual, seasonal, intra-seasonal, and synoptic), data were split in half because the length of the dataset should be at least double the length of the longest period considered in order to fit a complete cycle (Frescura et al. 2008). In this case the longest period considered in the partition of spectral densities is approximately 3 years (i.e. inter-annual variability), therefore, a minimum of 6 years (roughly half of the length of the full dataset) is required to perform the spectral analysis when the dataset is divided. The same analysis used for the complete datasets was applied to each of the two data blocks separately, and cumulative spectral densities were calculated for the same periods, partitioning the variance in SST and chl-a into their inter-annual, seasonal, intra-seasonal, and synoptic components.

### 2.1.2. Long term trends in environmental conditions

Seasonal trends on the south coast were calculated for SST, chl-a, water currents, and winds, for the same spatial area as in the previous section. The time span for the complete series was selected to start at the same time for all four variables, and covered approximately 15 years (from July 2002 until the beginning of 2017, see Table 1 for details on each variable). Given the need to have contemporaneous data, the period of analysis was limited by the availability of SST and chl-a. Although satellite measurements of SST and chl-a are available from the late 1970s, only MODIS Aqua was used to avoid inconsistencies in the series resulting from the different performances of multiple measuring instruments. The periods used to calculate the seasonal averages for the austral hemisphere followed the same time periods that appear in Brown (1992), i.e. June – August (winter), September – November (spring), December – February (summer), and March – May (autumn). In contrast to the spectral partitioning analysis where temporal resolution needs to be maximised at the expense of averaging data from larger areas, the analysis of seasonal trends involves seasonally averaged data, therefore minimising the impact of consecutive gaps in the data and allowing to maximise the spatial resolution for this analysis. Therefore, for each variable, seasonal averages were calculated for each year and area, using the highest spatial resolution possible. Although resolution was different for each variable, for simplicity, the smallest area used for each variable will be referred to as a *pixel*. The shortest temporal resolution available for each variable was also selected. In this case, eight-day averages, at 4km resolution were used for SST and chl-a. Water currents along the coast were obtained from the model Ocean Surface Current Analyses Real-time (OSCAR), for zonal and meridional current speeds predicted at 15m depth, using five-day averages with 0.33° resolution (https://coastwatch.pfeg.noaa.gov/erddap/griddap/jplOscar_LonPM180.html, ERDDAP server from NOAA). Monthly averages of wind data (ERA – Interim), were obtained from the European Centre for Medium-Range Weather Forecasts for zonal and meridional direction, as well as wind speed, measured at 10m height, with 0.125° resolution (data downloaded from https://apps.ecmwf.int/).

In the case of water currents and winds, the alongshore component (i.e. the west-east component), and the vector (calculated as the square root of the sum of squares of the zonal and meridional speed), were used in the regression analysis. To remove trends in the variability among years, seasonal anomalies were calculated for each variable as the difference between the seasonal average for the year and the seasonal average for the entire series. A simple linear regression of the magnitude of the seasonal anomaly versus year was calculated for each pixel and the magnitude and significance of the obtained slope was considered as a measurement of the long-term trend for each time series. The function *lm* from the package *stats* in R (R Core Team, 2015, version 3.2.0) was used. All figures were plotted using the package *ggplot2* (Wickham 2009).

2.2. Spatial patterns in mussel recruitment and adult cover

Ten sites were selected along the south coast of South Africa to estimate mussel recruitment and adult cover from Cape Agulhas to Morgan´s Bay (20° to 29° E, approximately, Fig. 1). Ten plastic pot scourers (referred to as *collectors*) were used as artificial substrata for mussel recruitment (Menge 1992, Porri et al. 2006a, von der Meden et al. 2010). Collectors are made from a tubular mesh that is rolled into a disc shape. These were embedded within the mussel bed during low tide, attached to eye-bolts. Recruitment was estimated for two entire months during each of two consecutive years: April and May of 2014 and 2015, corresponding with the autumn peak of recruitment (McQuaid and Lawrie 2005, McQuaid and Phillips 2006, Porri et al. 2006a, Porri et al. 2006b, Porri et al. 2008). Despite the period sampled was limited, sampling during the recruitment peak during both years reflected the longitudinal pattern of recruitment abundances along the coastline, and this pattern was consistent among months and years (Fig. S1), as well as with the adult mussel cover (Fig. 1D), which has been shown to be highly stable in time (Reaugh-Flower et al. 2011). Due to the large spatial extent of the area sampled, collectors were replaced only once per month during the same spring tide period, with the exception of Plettenberg in May 2014, where sampling was delayed for two weeks due to weather conditions. After one month of deployment, collectors were removed from the rock and preserved in 70% ethanol immediately after collection until samples were processed in the laboratory. For each month and site, three replicates were analysed when possible. Some collectors were detached from the rock by rough sea conditions, which limited the number of replicates for some sites (total n = 113). Mussel recruits were removed from the collectors by the addition of 10 ml of bleach in 250 ml of ethanol to dissolve the byssal threads (Connolly et al. 2001). Each collector was carefully unrolled and gently scrubbed in a bucket until all particles present were removed. Contents were filtered through a 75 μm sieve and stored in 70% ethanol. All individuals in each sample were counted except in the case of Brenton and Plettenberg in April and May of 2015. Due to the high abundances reached in these samples, only a 25% sub-sample was counted for those samples. Only the predominant mussel species were counted, identified, and divided in two groups: (1) *Perna perna,* and (2) other mytilids (comprising mostly *Mytilus galloprovincialis* and a few *Choromytilus meridionalis*). Adult cover at each site was estimated by photographic analysis by the point-intercept method using 49 evenly distributed points within each of 10 25x25 cm quadrats placed randomly in the mussel zone (Navarrete and Manzur 2008). Due to the high stability shown in percent cover of the mussel bed along the South African coastline during a five-year period (Reaugh-Flower et al. 2011), a single sampling was conducted in autumn 2015.

2.2.1. Statistical analysis

Most of the temporal variability in SST and chl-a corresponded to seasonal and intra-seasonal frequencies, and consequently, we focussed on the relationships of these two frequencies of variability with the spatial patterns of recruitment and adult cover of mussels. Simple linear regressions were performed on $\log_{10}$ 1+number of recruits for *Perna perna* and other mytilids separately, as a function of annual average chl-a and SST conditions, and seasonal and intra-seasonal cumulative spectral densities, estimated for the megapixel closest to each recruitment site (represented by squares in Figs. 1 B and C, 2 and 3). Logarithmic transformation was applied to make the relationships linear. Distance correlation (Székely and Rizzo 2009), and Pearson correlations were applied to all the models. Distance correlation allows the capture of non-linear relationships better than linear methods like Pearson correlation, and allows one to establish if variables are independent, providing a coefficient $R$ (between 0 and 1, where $R = 0$ for independent variables). This method was used to account for non-linearity and to compare the goodness of fit of linear vs. non-linear relationship, to support the linear model results. Linear regressions were run using the function *lm*, and Pearson correlations with the function *cor.test*, both from the package *stats*. Distance correlations were run with 999 permutations using the function *dcor.test*, from the package *energy* (Rizzo and Székely 2019). For each comparison, distance correlations were performed 10 times to obtain a range of *p*-values, a common procedure when applying permutations (e.g. Tapia et al. 2010).

3. Results

Environmental variables presented contrasting spatial distributions along the south coast of South Africa. Average values for the 15-year period showed the Agulhas Current as a warmer area (over 22°C within its core), especially to the northeast where the shelf is narrower and the current runs closer to the shore (Fig. 1C). The Agulhas Current inner border was clearly depicted as a sharp thermal front along the shelf break (shown as a marked change in colour) where temperatures dropped below 20°C. Over the wider western Agulhas Bank, temperatures were between 18 and 19°C, and slightly cooler onshore around Cape Agulhas. Primary productivity peaked nearshore, especially along the central Bank (22-25°E), reaching chl-a concentrations close to 7 mg m$^{-3}$ around the Plettenberg Bay area (Fig. 1B). The lowest chl-a concentrations (< 1 mg m$^{-3}$) characterised the Agulhas Current path offshelf, especially to the northeast, and the offshore western part of the Agulhas Bank.

Estimates of mean adult cover and recruitment rates of mussels peaked at Brenton and Plettenberg Bay along the central coast of the Agulhas Bank (Fig. 1D). This pattern was especially evident in the recruitment rates of *Perna perna* and other mytilids, with maximum values around $10^3$ individuals m$^{-2}$ day$^{-1}$. Adult cover surpassed 80% at these two central sites, although it was also relatively high to the west, above 40%, with the exception of Arniston. There was a decline in adult cover to the east, with estimates generally below 40% and even approaching 0% at Haga Haga, which, although less marked, followed the general spatial pattern observed in recruitment.

### 3.1. Temporal variability in environmental factors

Partitioning of spectral densities showed that most of the temporal variability in SST along the south coast of South Africa corresponded to seasonal cycles; these accounted for almost 75% of total temporal variability over the western Agulhas Bank (Fig. 2A). The remaining variability was explained by cycles longer than one year, which showed a very similar spatial pattern (Fig. S2 A). Shorter frequencies in the intra-seasonal range gained importance (explaining up to 35% of variability) along the shelf break, corresponding spatially with the inner border of the Agulhas Current and the upwelling cell at Port Alfred (27°E, Fig. 2B). Along this area, inter-annual variability accounted for approximately 15% of total variability, and there was a very small contribution by synoptic cycles (Fig. S2 A-B). This pattern of decreased seasonality and increased importance of intra-seasonal periods was especially evident during the second half of the study period (February 2010 and September 2017) along the shelf break (Fig. 2E and F). Thus, the reduction of seasonality from 50 to 35% was compensated by an equivalent increase in intra-seasonal cumulative spectral densities around the Port Alfred upwelling cell (Fig. 2). No marked change in inter-annual or synoptic cycles was evident (Fig. S2 C-F), except for a small decrease in inter-annual seasonality over the central and western Agulhas Bank (Fig. S2 C and E).

In contrast to SST, chl-a seasonal frequencies did not dominate throughout the whole region (Fig. 3). Intra-seasonal periods explained > 60% of total variability along the shelf break, while seasonal and intra-seasonal periods co-dominated over the central and western Agulhas Bank (Fig. 3A and B). During the second half of the study period there was a marked shift towards more variability being explained by intra-seasonal periods, especially over the central Agulhas Bank. In particular, around the Plettenberg Bay area, cumulative spectral densities in the seasonal range dropped from 50 to 30%, while an increase of similar magnitude was observed in the intra-seasonal range (Fig. 3C-F). Inter-annual cycles explained up to 15% of the variability over the western Agulhas Bank (Fig. S3 A), and changes between the two dataset halves were small (Fig. S3 C and E). Synoptic frequencies contributed to variability along the shelf break (Fig. S3 B), and a small increase in synoptic variability (approximately 10%) was detected in the western and central Agulhas Bank (Fig. S3 D and F).

### 3.2. Relationships between environmental and biological variables

Linear regression analysis showed that average chl-a and SST along the coast during the 15-year period of study, only explained around 30% of the variability in recruitment for the two mussel taxa (Fig. 4A and B). In contrast, temporal variability of chl-a was a very good predictor of spatial variability in mussel recruitment, with seasonal and intra-seasonal frequencies of variability explaining almost double the variance explained by average annual chl-a alone. Seasonal frequencies of variability in chl-a explained 67 and 73% of variance for *P. perna* and other mytilids respectively, while intra-seasonal frequencies explained 55 and 58% respectively (Fig. 4C and E). Recruitment for these taxa markedly increased with seasonal frequencies and decreased as intra-seasonal variability increased. On the other hand, temporal variability in SST was very weakly correlated with mussel recruitment rates. Seasonal frequencies of variability in SST only explained 17 and 15% of variance for *P. perna* and other mytilids respectively, and intra-seasonal frequencies explained 15 and 13% respectively (Fig. 4D and F). Strength of the relationship was very similar for both Pearson and distance correlation results, although the relationship improved when accounting for non-linearity in seasonal and intra-seasonal SST recruitment results (Table 2).

Similar results were obtained for patterns in adult cover as they correlated much better with temporal variability estimates of chl-a than with mean values for this variable (Fig. 5). Like recruitment rates, adult cover estimates increased with seasonal and decreased with intra-seasonal cumulative spectral densities in chl-a, each of which explained more than 60% of total variability (Fig. 5C and E), while the association with chl-a mean concentrations was not significant (Fig. 5A). In contrast, neither SST means nor their temporal components explained adult cover spatial patterns (Fig. 5B, D and F).

### 3.3. Long term trends in environmental conditions

Analysis of SST trends showed significant cooling during autumn, one of the main mussel larval recruitment periods (Fig. 6A). Coastal cooling at the Plettenberg (23°E) and Port Alfred (27°E) upwelling cells of around 0.1°C year$^{-1}$ was observed. Off Port Alfred significant SST declines were detected along the outer shelf to the south west following the inner border of the Agulhas current (Fig. 6A). Similar, although much weaker and non-significant, temperature drops were observed in spring and summer, with slight, non-significant warming of 0.05°C

year$^{-1}$ in winter (Fig. S4 A-D). Temporal changes in chl-a did not correspond to patterns in SST slopes as primary productivity levels remained largely constant, although slight non-significant reductions of around 0.25 mg m$^{-3}$ year$^{-1}$ were observed at both upwelling cells in autumn (Fig. S4 E-H). In autumn, a southwestwards acceleration of around 2cm s$^{-1}$ year$^{-1}$ of the Agulhas Current was observed along its path over the Eastern Agulhas Bank shelfbreak from approximately 27-24°E. On the other hand, eastwards currents increased by around 1 cm s$^{-1}$ year$^{-1}$ over the western Agulhas Bank over the same period. These patterns were also observed in spring, summer and winter (Fig. S4 I-L), but did not match the seasonal slopes found for zonal winds, which show non-significant increases in the magnitude of the westerlies, especially in winter (Fig. S4 M-P).

## 4. Discussion

Understanding population-level responses to directional change in environmental conditions is central to predicting how ecosystems will respond to ongoing climate change. In the case of the vast majority of benthic organisms, this is complicated by a biphasic life history and particularly important is the regulation of recruitment success. Our main results show that higher recruitment rates and estimates of adult abundances in foundational species in the form of coastal mussels were related more strongly to the prevailing temporal scale of fluctuation in chl-a than to the mean values of environmental variables. For marine species with biphasic life histories, the timing and success of reproduction can be explained through temperature relationships (e.g. Shanks et al. 2020; Émond et al. 2020), but they are often tightly linked to the availability of food for their planktotrophic larvae (Himmelman 1975; Starr et al. 1990; Giangrande et al. 1994; Gianasi et al. 2019). On a seasonal basis, this is relatively predictable and reflects seasonal changes in the conditions that control primary production, particularly light and mixing of the water column. Less predictable, intra-seasonal effects regulating food availability and the delivery of larvae to the adult habitat have direct effects at the population level. Here, we examined temporal variability in multiple environmental variables at four different temporal periods and found that the critical variable was chl-a. Both recruitment and adult cover showed a positive relationship with seasonal variability in chl-a, while exhibiting a negative relationship with intra-seasonal variability in chl-a.

Importantly, the temporal scales of variability in chl-a in our study area reflect large scale climatic forcing and the importance of teleconnections. Essentially, changes in the levels of seasonal and intra-seasonal variability in chl-a over the Eastern Agulhas bank are likely to be driven by recent changes in the behaviour of the Agulhas Current. An acceleration of the current (Rouault et al. 2009), as well as increased eddy activity (Beal and Elipot 2016) have been reported. The acceleration of the current appears to be driven by increasing wind forcing in the equatorial Indian Ocean (Rouault et al. 2009) and leads to enhanced coastal upwelling at the two upwelling cells of Plettenberg and Port Alfred. Similarly, eddy activity has also been described as a key mechanism influencing upwelling activity in the area (Goschen et al. 2015, Leber et al. 2017). Thus, changes in hydrodynamic processes driven by the Agulhas Current have the potential to affect coastal populations not primarily through changes in mean environmental conditions but mainly through changes in the temporal variability of conditions, particularly a weakening of seasonality in primary productivity.

Predictable, seasonal peaks in primary productivity can have multiple beneficial effects on mussel populations. Adult mussels can feed on particulate matter other than phytoplankton (Asmus and Asmus 1991, Puccinelli et al. 2016) so that chl-a is not a perfect proxy for preferred food, but strong seasonality improves the condition and physiological state of bivalves through the availability of high quality food (Newell and Bayne 1980, Norkko and Thrush 2006). While there is no reason to expect a direct link between reproductive output and recruitment because of larval advection, we would expect a relationship at the metapopulation level. High reproductive effort during the optimal environmental window in the year is expected in strongly seasonal systems (Varpe 2017) and enhanced investment in *a priori* allocation of energy for a single highly predictable annual reproductive peak has been observed in bivalves in such environments (Ćmiel et al. 2019). At the same time, improved recruitment with increased seasonality in chl-a reflects the fact that the survival and development of planktotrophic veliger larvae depend on the availability of phytoplankton (Raby et al. 1997, Vargas et al. 2006). Thus, clear seasonal signals in chl-a will favour both adult reproductive output and larval survival with obvious implications for the long-term sustainability of adult populations at the metapopulation level.

The critical implication is that evaluating patterns in recruitment and community composition of meroplanktonic intertidal organisms requires an understanding of the physical processes that influence the periodicity as well as the magnitude of change in environmental conditions. The study area experiences strongly seasonal reversals in the predominant wind direction with easterly winds in summer and winter westerlies (Schumann 1987) and this drives the upwelling of sub-surface cold water masses in the lee of headlands on this coast (Schumann et al. 1982). Analyses of latent heat flux have found that the sort of intra-seasonal variability that we observed is characteristic of strong sea surface temperature fronts at mid-latitudes, like the edges of the Gulf Stream and the Agulhas Current (Grodsky et al. 2009). Similarly, flow within the Eastern Australian Current was observed to vary on scales of 2-3 months (Creswell et al. 2017). We hypothesise that such temporal scales also characterise Agulhas Current driven processes like coastal upwelling induced by bottom friction at the shelfbreak (Gill and Schumann 1979, Fig. 6C) with winds driving upwelling on seasonal scales. Although both processes may be behind upwelling dynamics in our two upwelling areas, intra-seasonal Agulhas-driven upwelling is likely to be more important at Port Alfred where the shelf edge is narrower than at Plettenberg (Fig. 1). This is a key difference between the two upwelling cells, with potentially major consequences for the intertidal communities they support. Moreover, it is consistent with the spatial patterns in mussel recruitment and abundances found in this and other studies, with Plettenberg exhibiting the highest adult cover and recruitment rates (Bownes and McQuaid 2006, von der Meden et al. 2008). We suggest that it is this spatial variability in the characteristic time scales of upwelling development that produces an irregular, highly complex spatial pattern in the distribution of mussels in the region that is quite different from that predicted by the abundant centre hypothesis.

Both recruitment and adult cover increased with enhanced seasonality in chl-a availability and decreased as more of the total variability occurred at shorter, intra-seasonal time scales. Thus, our results indicate that it is important to include temporal variability in environmental conditions as well as their averages in models that aim to explain species distributions and community responses to long-term climate changes. Similarly, Baldanzi et al. (2015) found that thermal sensitivity of an intertidal sandhopper was related to temperature predictability as well as temperature variability. In another example, the strength and even the sign of the relationships between environmental variables and recruitment rates of mussels and barnacles along the Northeastern Pacific has varied repeatedly for more than 20 years (Menge et al. 2011). One unexplored possible explanation could be that the time scales at which these variables fluctuated also changed, leading to alternating weak/strong associations with recruitment.

The periodicity of environmental fluctuations can change with time and we observed an apparent homogenization of these scales towards greater intra-seasonal fluctuations in chlorophyll concentrations at the expense of seasonality, particularly in the Plettenberg upwelling cell. Thus, intra-seasonal Agulhas-driven upwelling is gaining importance at the expense of seasonal wind driven upwelling. This is consistent with upwelling intensification at the two cells detected by significant drops in SST, which has been observed in previous studies (Arnone et al. 2017). As no clear trends in local wind dynamics were observed, these changes seem to be driven by Agulhas Current driven topographic upwelling. This is supported by the observed acceleration of the Agulhas Current along the Eastern margin of the Agulhas Bank as this would strengthen this form of upwelling (Fig. 6B, Fig. S4). Moreover, the intensification of eastwards currents over the western Agulhas Bank is consistent with a reinforced cyclonic eddy driven by the Agulhas Current that is known to occur in that area (Harris 1978, Shannon and Chapman 1983). Simulations obtained through regional ocean models (ROMs) developed for this area (Rouault et al. 2010) indicate an increase in water transport, which is again consistent with acceleration of the Agulhas Current. Similarly, Backeberg et al. (2012) also reported an increase in eddy instability in the area adjacent to the east coast of South Africa, which is likely to influence our study area.

Western boundary currents, like the Agulhas Current, are forced by the wind conditions in their ocean basin, and strongly influence meteorological conditions in the adjacent areas (Lutjeharms 2006). Our data show no sign of significant changes in wind stress over our study region for the period examined (Fig. S4), but SST and wind conditions in the Western Indian Ocean vary among years, depending on variability in the Indian Ocean Dipole (Saji et al. 1999, Saji and Yamagata 2003). Rouault et al. (2009) reported a cooling trend around the upwelling cell at Port Alfred, which they ascribed to an increase in Agulhas flow driven by increases in wind forcing over the Western Indian Ocean basin. Wu et al. (2012) reported an intensification of the anticyclonic wind stress curl in the Indian Ocean gyre, and suggested that this could account for changes in Agulhas Current flow. This sequential mechanism creating teleconnections from a basin-scale climatic process to regional upwelling intensification may not only alter mean nearshore thermal and productivity mean conditions, but, more importantly, their seasonality, which affects intertidal populations more powerfully.

Long-term interannual variability in major current systems driven by anthropogenic climate change is known to affect the oceanography of adjacent coastal areas, although such trends seem to be specific to each system (Beal and Elipot, 2016). Mediation of coastal upwelling by variability in the circulation of the Kuroshio Current, the Gulf Stream and the Eastern Australian Current has been observed (McClain et al. 1984, Roughan and Middleton 2004, Xue et al. 2004). While the acceleration of the Kuroshio has been linked with historical cooling events during the Holocene (Zhang et al. 2019), the Eastern Australian Current has strengthened in response to recent global warming (Cetina-Heredia et al. 2014). In contrast, under the same global climatic forcing, the Gulf Stream has weakened (Dong et al. 2019). Such changes are likely to be translated into shifts in both the mean values and time scales of variability in environmental conditions across a wide range of coastal systems affected by these currents, although the potential consequences for coastal biological communities have not yet been extensively explored (but see Shanks et al. 2020). In addition, the dynamics of large scale currents may not be the only source of change in the phenology of primary productivity in nearshore waters. Seasonality of many phytoplankton groups along the Peruvian coast is known to weaken significantly during El Niño events (Espinoza-Morriberón et al. 2017). Similarly, both El Niño and the anomalous warm water mass in the North Pacific known as "The Blob" affected phytoplankton seasonality in Baja California (Jimenez-Quiroz et al. 2019). In the North Atlantic, changes in bloom magnitude, timing and periodicity were driven by a negative trend in the North Atlantic Oscillation during the first years of the 21$^{st}$ century (González Taboada and Anadón 2014). On a global scale, a conspicuous reduction in the seasonal amplitude of primary production has been forecasted by 6 different climatic models for the year 2100 due to poleward spreading of subtropical, less seasonal regimes driven by global warming (Henson et al. 2013). In this case, coastal populations of organisms adapted to marked seasonal fluctuations of their food source are likely to be profoundly negatively affected.

### 4.1. Conclusion

In conclusion, we found the scales of temporal fluctuation in primary productivity to show a strong relationship with recruitment and spatial patterns of adult mussel abundance at the metapopulation level, with seasonal, predictable fluctuations positively correlated with both these proxies of biological fitness. Over recent years, however, Agulhas Current mediated upwelling has been intensified by large scale climatic forcing, increasing the importance of intra-seasonal variability, with negative implications for mussel populations. Assessing possible responses of coastal ecosystems to climate change requires consideration of not only mean environmental conditions but also of the temporal scales at which they fluctuate.

**CRediT authorship contribution statement**
**Carlota Muñiz**: Conceptualization, Investigation, Formal analysis, Visualization, Writing – original draft. **Christopher D. McQuaid**: Conceptualization, Writing – original draft, Funding acquisition. **Nicolas Weidberg**: Conceptualization, Investigation, Visualization, Writing – original draft.

**Declaration of competing interest**

The authors declare that they have no known competing financial interests or personal relationships that could have appeared to influence the work reported in this paper.

**Acknowledgements**

This work is based upon research supported by the South African Research Chairs Initiative of the National Research Foundation of South Africa (Grant number 64801). During data processing and manuscript elaboration Nicolas Weidberg was funded by the NASA grant 80NSSC20K0074. We would also like to thank to all the volunteers who helped with fieldwork.